\documentclass{article}
\usepackage{spconf}
\pdfoutput=1
\usepackage{graphicx}
\usepackage{tikz}
\usepackage{amsmath}
\usepackage{amssymb,physics}
\usepackage{algorithm,setspace}
\usepackage{booktabs}
\usepackage{pifont}
\usepackage{tabularx,ragged2e}
\newcolumntype{Y}{>{\centering\arraybackslash}X}

\usepackage{pgfplots}
    \usetikzlibrary{calc,positioning,patterns}
    \pgfplotsset{compat=1.3}
\usepackage{pgfplotstable}
\usepackage[nolist, nohyperlinks]{acronym}
\usepackage{xfrac}
\usepackage{multirow}
\usepackage{comment}

\usepackage[subtle]{savetrees}
\newcommand{\ssp}[0]{\text{ }}

\newcommand{\D}[1]{\mathrm{d}{#1}}

\newcommand{\submin}[0]{_\text{min}}
\newcommand{\submax}[0]{_\text{max}}

\newcommand{\T}{\mathrm{T_{60}}}

\tikzstyle{block} = [draw, fill=blue!14, rectangle, minimum height=3em, minimum width=5em, align=center]
\tikzstyle{sum} = [draw, fill=white, circle, node distance=1cm]
\tikzstyle{input} = [coordinate]
\tikzstyle{output} = [coordinate]
\tikzstyle{pinstyle} = [pin edge={to-,thin,black}]
\tikzstyle{branch}=[fill,shape=circle,minimum size=3pt,inner sep=0pt]
\tikzstyle{connarrow}=[-latex, line width=1pt]
\tikzstyle{connline}=[-, line width=1pt]

\title{Analysing Diffusion-based Generative Approaches versus Discriminative Approaches for Speech Restoration}
\name{Jean-Marie Lemercier$^{\star}$, Julius Richter$^{\star}$, Simon Welker$^{\star \cross}$,
Timo Gerkmann$^{\star}$\thanks{This work has been funded by the Federal Ministry for Economic Affairs and Climate Action, project 01MK20012S, AP380, DASHH (Data Science in Hamburg - HELMHOLTZ Graduate School for the Structure of Matter) with the Grant-No. HIDSS-0002, and the German Research Foundation (DFG) in the transregio project Crossmodal Learning (TRR 169).}}
 
\address{$^{\star}$Signal Processing (SP), Universität Hamburg, Germany\\ 
	$^{\cross}$ Center for Free-Electron Laser Science, DESY, Germany\\
	{ \tt \small \{firstname.lastname\}@uni-hamburg.de}
	}
	
\begin{document}

\ninept

\begin{acronym}
\acro{stft}[STFT]{short-time Fourier transform}
\acro{istft}[iSTFT]{inverse short-time Fourier transform}
\acro{dnn}[DNN]{deep neural network}
\acro{pesq}[PESQ]{Perceptual Evaluation of Speech Quality}
\acro{polqa}[POLQA]{perceptual objectve listening quality analysis}
\acro{wpe}[WPE]{weighted prediction error}
\acro{psd}[PSD]{power spectral density}
\acro{rir}[RIR]{room impulse response}
\acro{snr}[SNR]{signal-to-noise ratio}
\acro{lstm}[LSTM]{long short-term memory}
\acro{polqa}[POLQA]{Perceptual Objectve Listening Quality Analysis}
\acro{sdr}[SDR]{signal-to-distortion ratio}
\acro{estoi}[ESTOI]{extended short-term objective intelligibility}
\acro{elr}[ELR]{early-to-late reverberation ratio}
\acro{tcn}[TCN]{temporal convolutional network}
\acro{rls}[RLS]{recursive least squares}
\acro{asr}[ASR]{automatic speech recognition}
\acro{ha}[HA]{hearing aid}
\acro{ci}[CI]{cochlear implant}
\acro{mac}[MAC]{multiply-and-accumulate}
\acro{vae}[VAE]{variational auto-encoder}
\acro{gan}[GAN]{generative adversarial network}
\acro{tf}[T-F]{time-frequency}
\acro{sde}[SDE]{stochastic differential equation}
\acro{drr}[DRR]{direct to reverberant ratio}
\acro{lsd}[LSD]{log spectral distance}
\acro{sisdr}[SI-SDR]{scale-invariant signal to distortion ratio}
\acro{mos}[MOS]{mean opinion score}
\end{acronym}

\maketitle
\begin{abstract}
Diffusion-based generative models have had a high impact on the computer vision and speech processing communities these past years. Besides data generation tasks, they have also been employed for data restoration tasks like speech enhancement and dereverberation. While discriminative models have traditionally been argued to be more powerful e.g. for speech enhancement, generative diffusion approaches have recently been shown to  narrow this performance gap considerably. 
In this paper, we systematically compare the performance of generative diffusion models and discriminative approaches on different speech restoration tasks.
For this, we extend our prior contributions on diffusion-based speech enhancement in the complex time-frequency domain
to the task of bandwith extension. 
We then compare it to a discriminatively trained neural network with the same network architecture on three restoration tasks, namely speech denoising, dereverberation and bandwidth extension. 
We observe that the generative approach performs globally better than its discriminative counterpart on all tasks, with the strongest benefit for non-additive distortion models, like in dereverberation and bandwidth extension.
Code and audio examples can be found online\footnote{https://uhh.de/inf-sp-sgmsemultitask}.

\end{abstract}

\section{Introduction}
\label{sec:intro}

Speech corruptions arise in real-life scenarios and modern communication devices, when clean speech sources are impacted by background noise, interfering speakers, room acoustics and channel degradation.
Speech restoration therefore aims at recovering clean speech from the corrupted signal.
Traditional speech restoration methods leverage the different statistical properties of the target and interference signals~\cite{gerkmann2018book_chapter}.
Data-driven approaches based on machine learning predominately employ discriminative models that learn a single best deterministic mapping between corrupted speech and the corresponding clean speech target~\cite{wang2018supervised}.

In contrast, generative models implicitly or explicitly learn the
target distribution and allow to generate multiple valid estimates
instead of a single best estimate as in discriminative approaches \cite{MurphyBook2}.
For example, diffusion-based generative models, or simply \emph{diffusion models}, have shown great success in learning the data distribution of natural images \cite{sohl2015deep,ho2020denoising,song2019generative}. 
This class of models uses a \textit{forward process} to slowly turn data into a tractable prior, such as a standard normal distribution, and train a neural network to solve the \textit{reverse process} to generate clean data from this prior.
These diffusion models can also be used for conditional generation in restoration tasks, which has recently been proposed for speech enhancement and dereverberation \cite{Welker2022SGMSE, Richter2022SGMSE++, lu2022conditional, serra2022universal}. 
They can in that regard be functionally seen as a mean of generating clean speech based on noisy speech, and can be thus compared to discriminative approaches. 
However, to make a fair comparison of these two conceptually different approaches, similar network architectures and same training data should be used.

In this work, we present an analysis of a generative diffusion model as compared to its discriminative counterpart sharing the same \ac{dnn} architecture, for various speech restoration tasks.
We use our previous method which defines the diffusion process in the complex spectrogram domain \cite{Welker2022SGMSE, Richter2022SGMSE++}.
We show that the performance gap between the generative and discriminative models varies with respect to the corruption at hand. 
We evaluate our proposed approaches on the WSJ0 corpus, using various simulated corruptions and recorded background noise. Finally we compare our bandwidth extension model with state-of-the-art bandwidth extension methods on the VCTK corpus.

The remainder of the paper is organized as follows. We first present the three speech restoration problems benchmarked, along with popular solutions for solving them. Then, we introduce diffusion-based generative models using the \ac{sde} formalism. We continue by explaining our experimental setup including data generation and training methods. Finally, we present and discuss our results.

\section{Speech Restoration Tasks and Related Work}
\label{sec:tasks}

\subsection{Speech enhancement}

Speech enhancement consists in removing an additive interference $n$ (e.g. background noise or interfering speakers) from the corrupted mixture $y$ to extract the clean speech target $s$:
\begin{equation}\label{eq:noise}
    y \ssp = \ssp s \ssp + \ssp n
\end{equation}

Popular enhancement methods include Wiener-inspired spectral filtering \cite{gerkmann2018book_chapter}, discriminative machine learning methods \cite{wang2018supervised}
or generative approaches like denoising \acp{vae} \cite{fang2021variational}.
Recently, diffusion models were proposed to tackle speech enhancement either in the time domain \cite{lu2021study}
or in the complex \ac{tf} domain \cite{Welker2022SGMSE, serra2022universal, Richter2022SGMSE++}.

\subsection{Speech dereverberation}

Reverberation is caused by room acoustics, and is characterized by multiple reflections on the room enclosures. Late reflections particularly degrade the speech signal and may result in reduced intelligibility \cite{Naylor2011}.
The corruption model is then convolutive, as the clean speech $s$ is convolved with a \ac{rir} $h$ representing the acoustic path between the source and the listener:
\begin{equation}\label{eq:reverb}
    y \ssp = \ssp s \ssp \ast \ssp h
\end{equation}

Single-channel dereverberation methods range from 
spectral enhancement \cite{Habets2007}, inverse filtering \cite{Kodrasi2014}, and cepstral processing \cite{Gerkmann2011} to machine learning algorithms using \acp{dnn} in the complex \ac{tf}
domain \cite{Williamson2017j}
or in the time-domain \cite{Ernst2019}.

\subsection{Bandwidth extension}

Audio super-resolution, or bandwidth extension, aims at converting a low-sampling rate signal back to a version sampled at a higher rate, regenerating time resolution, high-frequency content and audio quality.
The corruption process is linear and involves an anti-aliasing low-pass filter followed by a decimation operation:
\begin{equation}\label{eq:subsampling}
    y \ssp = \ssp \mathrm{Resample}(s \ssp \ast \ssp a, \ssp  ~f^\mathrm{up}_s, \ssp ~f^\mathrm{low}_s)    
\end{equation}
where $a$ is the anti-aliasing filter impulse response, $f^\mathrm{up}_s$ the original high sampling rate and $f^\mathrm{low}_s$ the low sampling rate.

Several discriminative methods were proposed to tackle bandwidth extension for speech signals \cite{
Birnbaum2017TFiLM, 
Nguyen2022Tunet}. Generative approaches based on neural vocoders using \acp{gan} were also proposed \cite{Liu2021VoiceFixer, Liu2022NeuralVI,Andreev2022Hifi++}.
A continuous-time diffusion model in the time-domain was proposed in \cite{Han2022NUWave2}.

\section{Score-based Diffusion Models for Speech Restoration} 
\label{sec:sgm}
Score-based diffusion models are defined by three  components: a forward diffusion process, a score estimator and a sampling method for inference.

\subsection{Forward and reverse processes}\label{sec:processes}

The stochastic forward process $\{\mathbf x_t\}_{t=0}^T$ is modeled as the solution to a \ac{sde}, in the Itô sense \cite{Oksendal2000SDE, song2021sde}: 
    \begin{equation} \label{eq:forward-sde}
        \D{\mathbf x_t} \ssp = \ssp \mathbf{f}(\mathbf x_t, ~t)\ssp  \D{t} \ssp + \ssp g(t) \ssp \D{\mathbf w}
    \end{equation}
where $\mathbf x_t$ is the current state of the process indexed by a continuous time variable $t \in [0, T]$ with the initial condition $\mathbf x_0$ representing clean speech. As our process is defined in the \ac{tf} domain, the variables in bold are assumed to be one-dimensional vectors in $\mathbb C^d$ containing the coefficients of a flattened complex spectrogram, whereas variables in regular font represent real scalar values. The stochastic process $\mathbf w$ is a standard $d$-dimensional Brownian motion, which implies that $\D{\mathbf w}$ is a zero-mean Gaussian random variable with standard deviation $\sqrt{\D{t}}$ for each \ac{tf} bin.

The \emph{drift} function $\mathbf f$ and \emph{diffusion} coefficient $g$ as well as the initial condition $\mathbf{x_0}$ and the final diffusion time $T$ define uniquely the Itô process $\{\mathbf x_t\}_{t=0}^T$.
Under some regularity conditions on $\mathbf f, g$ allowing a unique and smooth solution to the Kolmogorov equations associated to \eqref{eq:forward-sde}, the reverse process $\{\mathbf x_t\}_{t=T}^0$ is another diffusion process defined as the solution of a \ac{sde}, with the following form \cite{anderson1982reverse, song2021sde}:
\begin{equation}\label{eq:reverse-sde}
    \D{\mathbf x_t} \ssp = \ssp 
        \left[
            -\mathbf f(\mathbf x_t, ~t) \ssp + \ssp g(t)^2\nabla_{\mathbf x_t} \log \text{ } p_t(\mathbf x_t)
        \right] \ssp \D{t}
        \ssp + \ssp g(t)\ssp \D{\bar{\mathbf w}}\,,
\end{equation}
where $\D{\bar{\mathbf w}}$ is a $d$-dimensional Brownian motion for the time flowing in reverse and $\nabla_{\mathbf x_t} \log \text{ } p_t(\mathbf x_t)$ is the \emph{score function}, i.e. the gradient of the logarithm data distribution for the current process state $\mathbf x_t$.

Speech restoration tasks can be considered as conditional generation tasks, i.e. generation of clean speech $\mathbf{x_0}$ conditioned by the corrupted speech $\mathbf{y}$. 
In \cite{Welker2022SGMSE, Richter2022SGMSE++} we proposed to incorporate the conditioning directly into the diffusion process by defining the forward process as the solution to the following Ornstein-Uhlenbeck \ac{sde} \cite{Oksendal2000SDE}:
\begin{equation}\label{eq:ouve-sde}
    \D{\mathbf x_t} \ssp =
        \ssp \underbrace{\gamma \ssp (\mathbf y \ssp - \ssp \mathbf x_t)}_{:= \ssp \mathbf f(\mathbf x_t, \mathbf y)} \ssp \D{t}
        \ssp + \ssp \underbrace{\left[ \sigma\submin \left(\frac{\sigma\submax}{\sigma\submin}\right)^t \sqrt{2\log\left(\frac{\sigma\submax}{\sigma\submin}\right)} \ssp \right]}_{:= \ssp g(t)} \ssp \D{\mathbf w}\,,
\end{equation}
with $\gamma$ a \emph{stiffness} hyperparameter, and $\sigma\submin$ and $\sigma\submax$ two hyperparameters controlling the \emph{noise scheduling}, that is, the amount of Gaussian white noise injected at each timestep of the process. 

\begin{figure}
    \centering
    \includegraphics[width=0.9\columnwidth]{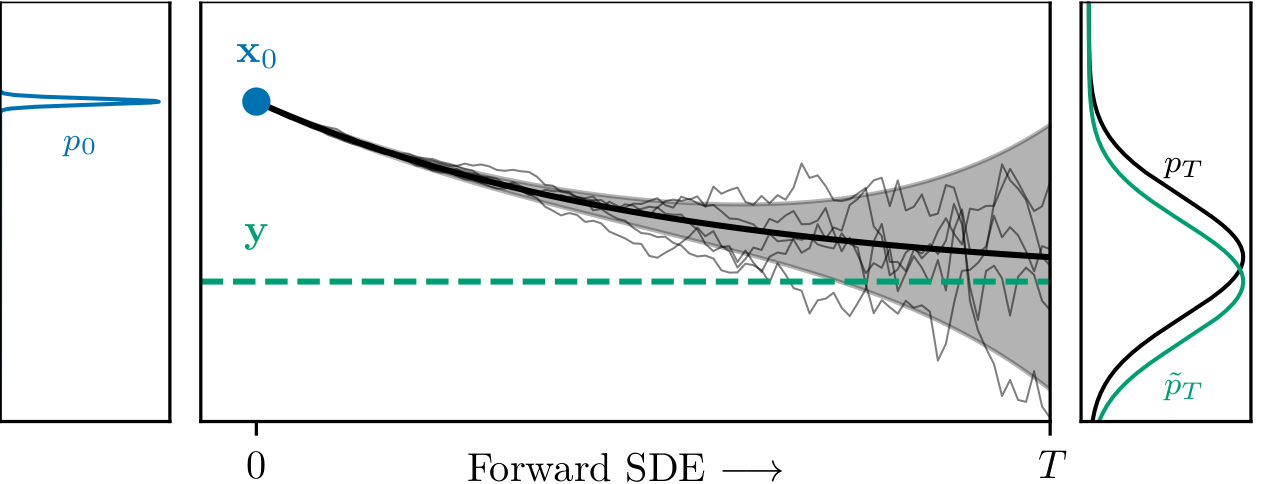}
    \vspace{-1em}
    \caption{\textit{Visualization of the forward process \eqref{eq:ouve-sde}. Mean curve is in solid black and variance is represented by the greyed area. Several realizations of the diffusion process are represented by thin black lines.}}
    \label{fig:process}
    \vspace{-1em}
\end{figure}

The interpretation of our forward process in Eq. \eqref{eq:ouve-sde}, visualized on Fig.~\ref{fig:process}, is as follows: at each time step and for each \ac{tf} bin independently, an infinitesimal amount of corruption is added to the current process state $\mathbf x_t$, along with Gaussian noise with standard deviation $g(t) \sqrt{\D{t}}$. 
Given an initial state $\mathbf x_0$ and $\mathbf y$, the Itô forward process corresponding to the solution of \eqref{eq:ouve-sde} admits a Gaussian distribution for the process state $\mathbf x_t$ called \emph{perturbation kernel}:
\begin{equation}
\label{eq:perturbation-kernel}
    p_{0t}(\mathbf x_t| \ssp \mathbf x_0, \ssp \mathbf y) \ssp = \ssp  \mathcal{N}_\mathbb{C}\left(\mathbf x_t; \ssp \boldsymbol \mu(\mathbf x_0, \ssp \mathbf y, \ssp t), \ssp \sigma(t)^2 \mathbf{I} \right),
\end{equation}
where $\mathcal{N}_\mathbb{C}$ denotes the circularly-symmetric complex normal distribution and  $\mathbf I$ the identity matrix. 
Given the simple Gaussian kernel, closed-form solutions for the mean $\boldsymbol \mu$ and variance $\sigma(t)^2$ can be determined \cite{Oksendal2000SDE}:
\begin{equation}
\label{eq:mean}
    \boldsymbol\mu(\mathbf x_0, ~\mathbf y, ~t) \ssp = \ssp \mathrm e^{-\gamma t} \mathbf x_0 \ssp + \ssp (1-\mathrm e^{-\gamma t}) \mathbf y
    \,,
\end{equation}%
\begin{equation}
    \label{eq:std}
    \sigma(t)^2 = \frac{
        \sigma\submin^2\left(\left(\sfrac{\sigma\submax}{\sigma\submin}\right)^{2t} \ssp - \ssp \mathrm e^{-2\gamma t}\right)\log(\sfrac{\sigma\submax}{\sigma\submin})
    }{\gamma \ssp + \ssp \log(\sfrac{\sigma\submax}{\sigma\submin})}
    \,.
\end{equation}
\vspace{-1em}

\subsection{Score function estimator}\label{sec:training}

When performing inference by sampling through the reverse \ac{sde} in Eq. \eqref{eq:reverse-sde}, the score function $\nabla_{\mathbf x_t} \log  \text{ } p_t(\mathbf x_t)$ is not readily available. Thus, it is approximated  by a DNN $\mathbf{s}_\theta$, called the \emph{score model}. In particular, given the Gaussian form of the perturbation kernel $p_{0t}(\mathbf x_t|\mathbf x_0, \mathbf y)$ and the regularity conditions exhibited by the mean and variance, a \emph{denoising score matching} objective can be used to train the score model $\mathbf{s}_\theta$ \cite{hyvarinen2005estimation}.

The score function of the perturbation kernel is:
\begin{equation}
    \nabla_{\mathbf x_t} \log \ssp p_{0t}(\mathbf x_t| \ssp \mathbf x_0, \ssp \mathbf y) \ssp = \ssp -\frac{\mathbf x_t \ssp - \ssp  \boldsymbol\mu(\mathbf x_0, \ssp \mathbf y, \ssp t)}{\sigma(t)^2}.
\end{equation}
Therefore we can reparameterize the denoising score matching objective as follows \cite{song2021sde}:
\begin{align}\label{eq:training-loss}
    \mathcal{J}(\theta)
    &= \mathbb{E}_{t, \mathbf x_0,\mathbf y, \{ \mathbf x_t|\mathbf x_0,\mathbf y \} } \left[
        \norm{\mathbf s_\theta(\mathbf x_t, ~\mathbf y, ~t) \ssp - \ssp \nabla_{\mathbf x_t} \log ~p_{0t}(\mathbf x_t| ~\mathbf x_0, ~\mathbf y) }_2^2
    \right] \nonumber \\
    &= \mathbb{E}_{t,\mathbf x_0,\mathbf y, \mathbf z} \left[
        \norm{\mathbf s_\theta \big( \big[ \boldsymbol\mu(\mathbf x_0 , ~\mathbf y, ~t \big) + \sigma(t) \mathbf z \big] , ~\mathbf y, ~t) + \frac{\mathbf z}{\sigma(t)}}_2^2
    \right],
\end{align}
using $\mathbf x_t =  \boldsymbol\mu(\mathbf x_0, \mathbf y, t) + \sigma(t) \mathbf z$,  with $\mathbf z \sim \mathcal{N}_\mathbb{C}\left(\mathbf z; ~\mathbf 0,  ~\mathbf{I}\right)$. $t$ is sampled uniformly in $[t_\epsilon, ~T]$ where $t_\epsilon$ is a minimal diffusion time used to avoid numerical instabilities.

\subsection{Inference through reverse sampling}\label{sec:inference}

At inference time, we first sample an initial condition of the reverse process, corresponding to $\mathbf x_T$, with:
\begin{equation}
    \mathbf x_T \ssp \sim \ssp \mathcal N_{\mathbb C}(\mathbf x_T; ~\mathbf y, ~\sigma^2(T) \mathbf I), 
\end{equation}
This sample corresponds to corrupted speech $\mathbf y$ to which we add Gaussian noise with variance $\sigma(t)^2$, which approximates the training condition.

Conditional generation is then performed by solving the following \emph{plug-in reverse \ac{sde}} from $t=T$ to $t=0$, where the score function is replaced by its estimator $\mathbf s_\theta$, assuming the latter was trained e.g. according to Section~\ref{sec:training}:
\begin{equation}\label{eq:plug-in-reverse-sde}
    \D{\mathbf x_t} \ssp =
        \ssp \left[
            -\mathbf f(\mathbf x_t, \ssp \mathbf y) \ssp + \ssp g(t)^2 \ssp \mathbf s_\theta(\mathbf x_t, \ssp \mathbf y, \ssp t)
        \right] \ssp \D{t}
        \ssp + \ssp g(t) \ssp \D{\bar{\mathbf w}},
\end{equation}
where $\mathbf{f}$ and $g$ are the drift and diffusion terms defined in \eqref{eq:ouve-sde}.

We use classical numerical solvers
based on discretization of
\eqref{eq:plug-in-reverse-sde} according to a $N$ points grid of the interval $[0, T]$.
Since each reverse diffusion step calls the score network, the inference time of diffusion models is higher than their discriminative counterparts, by two orders of magnitude in our case. Fast inference schemes are discussed in the literature and are outside of the scope of this paper.

\vspace{-0.25em}

\section{Experimental Setup}
\label{sec:exp}

\newcommand{\parjump}{0.3em}
\vspace{-1em}

\subsection{Data}

We use the WSJ0 corpus \cite{datasetWSJ0} 
for most experiments to ensure easier comparison between tasks. For comparison to bandwidth extension baselines, we use the VCTK corpus \cite{valentini2016investigating}. All data generation methods are accessible via our web page\footnote{https://uhh.de/inf-sp-sgmsemultitask}.

\textit{Speech Enhancement}: \,The WSJ0+Chime dataset is generated using clean speech extracts from the Wall Street Journal corpus
and noise signals from the CHiME3 dataset \cite{barker2015third}. The mixture signal is created by randomly selecting a noise file and adding it to a clean utterance with a \ac{snr} sampled uniformly between 0 and 20$\,$dB.
\vspace{\parjump}

\textit{Speech Dereverberation}: \,The WSJ0+Reverb dataset is generated using clean speech data from the WSJ0 dataset and convolving each utterance with a simulated \ac{rir}. We use the PyRoomAcoustics
engine \cite{Scheibler2018PyRoom} to simulate the \acp{rir}. The reverberant room is modeled by sampling uniformly a target $\T$ between 0.4 and 1.0 seconds and room length, width and height in [5,15]$\times$[5,15]$\times$[2,6] m. A dry version of the room is created with the same geometric parameters with a fixed absorption coefficient of 0.99, to generate the corresponding anechoic target. 
\vspace{\parjump}

\textit{Bandwidth Extension}: \,The WSJ0+BWR dataset is built with clean speech extracted from the WSJ0 corpus and a similar bandwidth reduction recipe as in \cite{Liu2021VoiceFixer, Andreev2022Hifi++}. We pick an anti-aliasing filter type among Chebyshev, Butterworth, Elliptic and Bessel and a filter order among \{2,4,8\}. Decimating is then realized with a down-scaling factor sampled in \{2,4,8\}. The utterance is then resampled at the original 16 kHz with polyphase filtering. To compare against other baselines, we generate VCTK+BWR by replacing WSJ0 with VCTK 
as the base speech corpus, which we first resample to 16kHz, and use the same process as explained above.

\vspace{-0.5em}
\subsection{Hyperparameters and training configuration}

\hspace{\parindent}\textit{Data representation}: \,Utterances are transformed using a \ac{stft} with a window size of 510, a hop length of 128 and a Hann window. Square-root magnitude compression is carried on the spectrogram.
For training, sequences of 256 \ac{stft} frames (i.e. 2s) are randomly extracted from the full-length utterances and normalized with respect to the corrupted mixture before being fed to the network. 
\vspace{\parjump}

\textit{Forward diffusion}: \,Defined in \eqref{eq:ouve-sde}, the stiffness parameter is fixed to $\gamma=$1.5, the extremal noise levels to $\sigma_\mathrm{min}=$~0.05 and $\sigma_\mathrm{max}=$~0.5. The minimal diffusion time defined in \eqref{eq:training-loss} is set to $t_\epsilon=$~0.03 as in \cite{Richter2022SGMSE++}. 
\vspace{\parjump}

\textit{Network architecture}: \,The original architecture used for score estimation in \cite{Richter2022SGMSE++} is the NCSN++ network proposed in \cite{song2021sde}. NCSN++ is a multiresolution U-Net structure which includes in each layer a series of ResNet blocks using 2D convolutions, group normalization and fixed down/upsampling.
Attention mechanism is used in the bottleneck, and the network leverages a parallel progressive growing path in addition to the skip connections.
The noisy speech spectrogram $\mathbf y$ and the current diffusion process estimate $\mathbf x_t$ real and imaginary channels are stacked and fed to the network as input. 
The model is made noise-conditional by feeding each ResNet block with an encoded version of the current noise level $\sigma(t)$
.
More details about the architecture can be found in \cite{Richter2022SGMSE++, song2021sde}.
For the generative model proposed in this paper, denoted as \textit{SGMSE+M}, we use a lighter configuration of the NCSN++ architecture called NCSN++M
.
Ablation studies were designed to halve the number of parameters with almost no degradation, resulting in a network capacity of roughly 27.8M parameters.
For the discriminative approach, denoted simply as \textit{NCSN++M} in the following, the noise-conditioning layers are removed
This ablation removes only 1.8\% of the original number of parameters, which hardly modifies the network capacity.
\vspace{\parjump}

\textit{Training configuration}: \,We train the \ac{dnn} for a maximum of 300 epochs using early stopping 
with a patience of 10 epochs%
. The generative approach SGMSE+M is trained with the denoising score matching criterion \eqref{eq:training-loss}, and discriminative NCSN++M uses a simple mean-square error loss on the complex spectrogram. We use the Adam optimizer
with a learning rate of $10^{-4}$ and an effective batch size of 16. We track an exponential moving average of the DNN weights with a decay of 0.999.
\vspace{\parjump}

\textit{Inference}: \,50 time steps are used for reverse inference, adopting the predictor-corrector scheme \cite{song2021sde} with one step of annealed Langevin dynamics correction.
\vspace{-0.75em}

\subsection{Evaluation metrics}

For instrumental evaluation of the speech enhancement and dereverberation performance, we use \ac{pesq} \cite{Rix2001PESQ}
, \ac{estoi} \cite{Jensen2016ESTOI}
and \ac{sisdr} \cite{Leroux2019SISDR}
. 
For bandwidth extension we also include \ac{lsd} as a common metric used in the literature. 
However, it must be stated that the aforementioned instrumental metrics may relate poorly with listening experiments, especially for bandwidth extension. We therefore complement our metrics benchmark with WV-MOS \cite{Andreev2022Hifi++}\footnote{https://github.com/AndreevP/wvmos}, which is a DNN-based \ac{mos} estimation, and was used by the authors for assessment of bandwidth extension performance. 
For comparability purposes to baselines on the VCTK corpus, we use regular STOI \cite{Thaal2011STOI} instead of its extended version
.

\vspace{-0.25em}

\section{Experimental Results and Discussion}
\label{sec:results}

\subsection{Speech enhancement}

In Table~\ref{tab:results:wsj0+chime}, we report speech enhancement performance on the WSJ0+Chime dataset. 
We notice that the generative SGMSE+M produces higher quality samples as measured by WV-MOS and PESQ. It is however slightly outperformed by discriminative NCSN++M on intelligibility and noise removal. 
Indeed, in a denoising task, the interference does not share any information with the target speech, making it relatively easy for a discriminative approach to remove the interference without distorting the target. However, we show in the uploaded listening examples that the discriminative approach tends to destroy low-energy speech regions for low \acp{snr}, whereas the generative model does not.
A larger benefit of the generative approach is observed when training and testing data have a stronger mismatch \cite{Richter2022SGMSE++}.

\begin{table}[t]
    \centering
    \caption{Results for denoising on WSJ0+Chime data.}
    \vspace{0.3em}
    \scalebox{0.8}{
    \begin{tabular}{c|c|cccc}
    
\toprule 
Method & Type & WV-MOS & PESQ & ESTOI & SI-SDR \\

\midrule
\midrule
Mixture & &
1.44 $\pm$ 1.62 & 1.70 $\pm$ 0.49 & 0.78 $\pm$ 0.14 & 10.0 $\pm$ 5.7 \\
\midrule
NCSN++M & D &
3.65 $\pm$ 0.48 & 2.67 $\pm$ 0.69 & \textbf{0.93 $\pm$ 0.06} & \textbf{19.5 $\pm$ 4.4} \\
SGMSE+M & G &
\textbf{3.77 $\pm$ 0.32} & \textbf{2.94 $\pm$ 0.60} & 0.92 $\pm$ 0.06 & 18.0 $\pm$ 5.1 \\
 \midrule
    \bottomrule
    \end{tabular}}
    \label{tab:results:wsj0+chime}
\vspace{-1.2em}
\end{table}

\subsection{Speech dereverberation}

In Table~\ref{tab:results:wsj0+rev}, we report dereverberation results on the WSJ0+Reverb dataset. 
Here, generative SGMSE+M clearly outperforms discriminative NCSN++M in terms of quality by a large margin on WV-MOS and PESQ, and performs on par on ESTOI and SI-SDR. 
For dereverberation, in contrast to denoising, the interference model is completely dependent on the target as it is a filtered version of the latter (Eq.~\eqref{eq:reverb}). The generative model is able to extract the speech cues and directly reconstructs it with very little reverberation. The discriminative method, however, cannot do so without introducing significant distortions.

\begin{table}[t]
    \centering
    \caption{Results for dereverberation on WSJ0+Reverb data.}
    \vspace{0.3em}
    \scalebox{0.8}{
    \begin{tabular}{c|c|cccc}
\toprule 
Method & Type & WV-MOS & PESQ & ESTOI & SI-SDR \\

\midrule
\midrule
Mixture &  &
1.78 $\pm$ 0.99 & 1.36 $\pm$ 0.19 & 0.46 $\pm$ 0.12 & -7.3 $\pm$ 5.5 \\
\midrule
NCSN++M & D &
2.96 $\pm$ 0.38 & 2.19 $\pm$ 0.48 & \textbf{0.87 $\pm$ 0.05} & \textbf{7.2 $\pm$ 3.7} \\
SGMSE+M & G &
\textbf{3.43 $\pm$ 0.33} & \textbf{2.64 $\pm$ 0.42} & \textbf{0.87 $\pm$ 0.05} & 6.4 $\pm$ 4.2 \\
 \midrule
    \bottomrule
    \end{tabular}}
    \label{tab:results:wsj0+rev}
\vspace{-1.2em}
\end{table}

\subsection{Bandwidth extension}

\begin{table}[t]
    \centering
    \caption{Results for bandwidth extension on WSJ0+BWR data.} 
    \vspace{0.5em}
    \scalebox{0.9}{
    \begin{tabular}{c|c|ccccc}
    
\toprule 
Method & Type & WV-MOS $\uparrow$ & ESTOI $\uparrow$ & LSD $\downarrow$ \\

\midrule
\midrule
Mixture & &  
2.45 $\pm$ 1.01 & 0.72 $\pm$ 0.21 & 2.31 $\pm$ 0.32 \\
\midrule
AE-NCSN++M & D &
 2.17 $\pm$ 0.93 & 0.71 $\pm$ 0.19 & 1.81 $\pm$ 0.21 \\
 
NCSN++M & D &
 2.25 $\pm$ 0.87 & 0.73 $\pm$ 0.16 & 2.21 $\pm$ 0.30 \\
 
SGMSE+M & G &
 \textbf{3.43 $\pm$ 0.48} & \textbf{0.83 $\pm$ 0.13} & \textbf{1.44 $\pm$ 0.17} \\
 
 \midrule
    \bottomrule
    \end{tabular}}
    \label{tab:results:wsj0+bwe}
\vspace{-1.2em}
\end{table}

\hspace{\parindent}\textit{Results on WSJ0+BWR}: \, In Table~\ref{tab:results:wsj0+bwe} we report bandwidth extension performance on the WSJ0+BWR dataset. 
Interestingly, using a STFT representation did not allow the discriminative approach to recreate the lost high-frequency content. The approach simply learnt an identity mapping, and similar results were observed when experimenting with other STFT-based DNN backbones and data. 
For this discriminative case, we modified the NCSN++M architecture to use a learnt encoder and decoder, as in e.g.
\cite{Birnbaum2017TFiLM}
. The resulting approach, denoted as AE-NCSN++M in the following, uses a single 1D convolutional layer with 256 filters of length 510 and stride 128, so that the learnt representation is equivalent to the chosen STFT filterbank.
As opposed to NCSN++M, AE-NCSN++M is able to generate high-frequency components, however the reconstruction quality is overall poor, which is to be expected given the generative nature of the bandwidth extension task. By learning an approximate identity mapping, NCSN++M performs better  than AE-NCSN++M on instrumental metrics although it does not actually perform bandwidth extension.

In contrast, generative SGMSE+M performs much better in all metrics, generating plausible content even when the Nyquist frequency is down to $2$kHz. 
 When the Nyquist frequency is down to $1$kHz, the approach can struggle with generating the right consonants in some cases. Typically the generation process may mistake $[ch]$ for $[s]$ as the information needed to differentiate those sounds is available only at frequencies way above $1$kHz. Integrating a linguistic or visual model could be here envisaged to make the approach robust to this lack of acoustic cues.
 \vspace{0.5em}

\textit{Results on VCTK+BWR}: \,In Table~\ref{tab:results:vctk-bwe16}, we compare the proposed generative SGMSE+M on the VCTK+BWR test set against HiFI++ \cite{Andreev2022Hifi++}, VoiceFixer\footnote{https://github.com/haoheliu/voicefixer} \cite{Liu2021VoiceFixer}, TFiLM \cite{Birnbaum2017TFiLM}, TUNet\footnote{https://github.com/NXTProduct/TUNet} \cite{Nguyen2022Tunet} and NuWave2\footnote{https://github.com/mindslab-ai/nuwave2} \cite{Han2022NUWave2}.

Please note that HiFi++, TFiLM and TUNet are trained on each input bandwidth separately, while our generative model SGMSE+M as well as VoiceFixer and NuWave2 are bandwidth-agnostic. 
We use the official implementations for all approaches without retraining,  except HiFi++ as no code is available, and TFiLM as no multi-speaker model is provided. When a method is not trained to restore speech at 16kHz, we use it at the nominal sampling rate then downsample its output to 16kHz. SGMSE+M achieves on par results with HiFi++ on 4kHz bandwidth and worse on 1kHz and 2kHz bandwidths, which is partially due to the fact that HiFi++ is trained separately for each input bandwidth. Using neural vocoders incorporating speech knowledge, as is the case for HiFi++, also probably helps improve robustness to very low input bandwidths.
Against all other approaches than HiFi++, SGMSE+M performs significantly better on almost all metrics and conditions.

\begin{table}[t]
    \centering
    \caption{Results for bandwidth extension on VCTK data. $^\star$~means that the results were taken from \cite{Andreev2022Hifi++}. $^\dagger$~means that the method was trained on each bandwidth reduction factor separately.}
    \vspace{0.25em}
    \scalebox{0.73}{
    \begin{tabular}{c|c|cc|cc|cc}
    
\toprule 
Bandwidth & Type & \multicolumn{2}{c}{1kHz} & \multicolumn{2}{c}{2kHz} & \multicolumn{2}{c}{4kHz} \\

 & & WV-MOS & STOI & WV-MOS & STOI & WV-MOS & STOI \\
\midrule
\midrule
Mixture & & 
 1.36 & 0.79 & 2.34 & 0.89 & 3.52 & 0.99 \\
 
\midrule

$\text{TUNet}^\dagger$ \cite{Nguyen2022Tunet} & D &
 - & - & - & - & 3.86 & 0.98 \\
 
$\text{TFiLM}^{\star\dagger}$ \cite{Birnbaum2017TFiLM} & D &
 1.65 & 0.81 & 2.27 & 0.91 & 3.49 & \textbf{1.00} \\
 
$\text{HiFi++}^{\star\dagger}$ \cite{Andreev2022Hifi++} & G & 
\textbf{3.71} & \textbf{0.86} & \textbf{3.95} & \textbf{0.94} & 4.16 & \textbf{1.00} \\

VoiceFixer \cite{Liu2021VoiceFixer} & G &
 2.50 & 0.73 & 3.35 & 0.78 & 3.81 & 0.83  \\

NuWave2 \cite{Han2022NUWave2} & G &
 - & - & - & - & 3.76 & 0.97 \\
 
SGMSE+M & G &
 3.25 & 0.83 & 3.70 & 0.93 & \textbf{4.20} & \textbf{1.00} \\
\midrule 
\bottomrule
    \end{tabular}}
    \label{tab:results:vctk-bwe16}
\vspace{-1.2em}
\end{table}

\vspace{-1em}

\section{Conclusion}
\label{sec:conclusion}
The goal of this work is to analyse the potential benefit of recent diffusion-based generative approaches against discriminative approaches on various speech restoration tasks.
For this, we apply our recently proposed diffusion generative model to speech enhancement, dereverberation and bandwidth extension, and compare against a discriminative approach using the same \ac{dnn} architecture
.
We observe that the generative approach performs globally better than its discriminative counterpart on all tasks, with the strongest benefit for non-additive distortion models, like in dereverberation and bandwidth extension.
Furthermore, we show that the proposed bandwidth-agnostic method performs slightly worse or on par in comparison with a recent bandwidth-dependent approach, and largely outperforms other discriminative and bandwidth-agnostic generative approaches.

\bibliographystyle{ieeetr}
\bibliography{biblio}

\end{document}